\documentclass[namedreferences]{solarphysics}

\usepackage[hyperref,optionalrh,showbiblabels]{spr-sola-addons} 
\usepackage{graphicx}        
\usepackage{color}           
\usepackage{breakurl}        

\newcommand{\aap}{    {\it Astron. Astrophys.}}

\newcommand{\apj}{    {\it Astrophys. J.}}
\newcommand{\apjl}{   {\it Astrophys. J. Lett.}}
 
\newcommand{\cjaa}{   {\it Chin. J. Astron. Astrophys.}} 

\newcommand{\grl}{    {\it Geophys. Res. Lett.}}

\newcommand{\jastp}{  {\it J. Atmos. Solar-Terr. Phys.}} 
\newcommand{\jgr}{    {\it J. Geophys. Res.}}
\newcommand{\mnras}{  {\it Mon. Not. Roy. Astron. Soc.}}

\newcommand{\solphys}{{\it Solar Phys.}}
 
\newcommand{\ssr}{    {\it Space Sci. Rev.}} 
\newcommand{\na}{    {\it New Astronomy}} 
\newcommand{\raa}{ {\it Research in Astronomy and Astrophysics}}
\newcommand{\pss}{    {\it Planet. Space Sci.}} 
\newcommand{\jcrl}{    {\it J.Commun. Res. Lab.}}
\newcommand{\nar}{    {\it New Astronomy Reviews }}
\newcommand{\sci}{     {\it Science}}
\newcommand{\livr} {   {\it Living Rev. Solar Phys.} }
\chardef\us=`\_

\begin{document}

\begin{article}
\begin{opening}

\title{The Broken Lane of a Type II Radio Burst Caused by Collision of a Coronal  Shock with a Flare Current Sheet \\ {\it Solar Physics}}

\author[addressref={aff1,aff2,aff3},corref,email={ggn@ynao.ac.cn}]{\inits{G.}\fnm{Guannan}~\lnm{Gao}}
\author[addressref=aff1]{\inits{M.}\fnm{Min}~\lnm{Wang}}
\author[addressref=aff4]{\inits{M.}\fnm{Ning}~\lnm{Wu}}
\author[addressref=aff1]{\inits{J.}\fnm{Jun}~\lnm{Lin}}
\author[addressref=aff5]{\inits{J.}\fnm{E.}~\lnm{Ebenezer}}
\author[addressref=aff3]{\inits{B.}\fnm{Baolin}~\lnm{Tan}}

\address[id=aff1]{Yunnan  Observatories, Chinese Academy of Sciences, Kunming, Yunnan 650011, China}
\address[id=aff2]{Key Laboratory for the Structure and Evolution of Celestial Objects, Yunnan Observatories, Chinese Academy of Sciences, Kunming, Yunnan 650011, China}
\address[id=aff3]{Key Laboratory of Solar Activity, National Astronomical Observatories, Chinese Academy of Sciences, Beijing 100012, China}
\address[id=aff4]{School of Tourism and Geography, Yunnan Normal University, Kunming, Yunnan 650031, China}
\address[id=aff5]{Indian Institute of Astrophysics Koramangala Bangalore 560034, India}

\runningauthor{Gao \emph{et al.}}
\runningtitle{The broken lane of a type II burst}

\begin{abstract}
We investigated  a peculiar  metric type II solar radio burst with  a broken lane structure, which was observed on November 13, 2012.  In addition to the radio data, we also studied the data in the other wavelengths.  The bursts were associated with two CMEs and two flares that originated from active region AR 11613. A long current sheet was  developed in the  first CME, and the second CME collided with the current sheet first and then merged with the first one. Combing information revealed by the multi-wavelength  data indicated that  a coronal shock accounting for the type II radio burst,  and that the collision of this shock with the current sheet resulted in the broken lane of the type II radio burst. The type II burst lane resumed after the shock passed through the current sheet. We further estimated the thickness of the current sheet according to the gap on the lane of the type II burst, and found that the result is consistent with previous ones obtained for various events observed in different wavelengths  by different instruments. In addition, the regular type II burst associated with the first CME/flare was also studied, and the magnetic field in each source region of the two type II bursts was further deduced in different way.
\end{abstract}
\end{opening}

\section{Introduction}
     \label{S-Introduction} 
     
A solar eruption is associated with a disruption of the coronal magnetic field,  in which the closed magnetic field in the low corona is severely stretched, and a magnetically neutral region, also known as the current sheet (CS), forms separating regions of oppositely directed magnetic fields. Magnetic reconnection takes place inside the CS at a reasonably fast rate, produces the solar flare  in the low solar atmosphere, and helps  the upper part of the erupting magnetic structure to escape to the outer corona and  interplanetary space, giving rise to a coronal mass ejection (CME). The CS is separating regions of oppositely directed magnetic fields
\citep[\textit{e.g.}, see also][]{for00,lin00,lin02,lin03,for06}.
     
In the case when the eruption is energetic enough, it produces a fast expanding CME that may further generate a coronal wave in front by the CME.  Coronal waves are considered to be signatures of large-amplitude fast-mode waves or shocks. These signatures are easily seen in EUV and SXR, and even in white light.
 \citep[\textit{e.g.}, see][]{vou03,ma11,kwo13}. Furthermore, type II bursts are the signatures of shocks traveling through the solar corona, so they give unambiguous evidence for fast mode shocks driven by the CME as opposed to other large-amplitude disturbances.
In dynamic spectra, a type II burst is often identified with two narrow parallel lanes in the metric to kilometric wavelength ranges. The lane of the lower frequency results from the fundamental band, and that of the higher energy from the harmonic band.  They  are produced by Langmuir turbulence in the plasma that is excited by the fast mode shock at the local electron plasma frequency, $f_{p}$. So the observed emission frequency $f_{obs}$ is related to $f_{p}$ and the electron density $n_{e}$ in the burst source region by $f_{obs}=s f_{p}, ~~f_{p}~[kHz] = ~8.98 \sqrt{n_{e}~[{cm}^{-3}]}$, and thus to the height of the source region if a coronal density model, $n_{e}$=$n_{e}(h)$, is given. Here $s$ is for the fundamental ($s=1$) and for the harmonic ($s=2$) band, respectively \citep{mcl85}. 

 There are two main classes of mechanisms for invoking type II radio bursts including  the CME-driven fast mode shock   and the blast wave \citep[\textit{e.g.},][and references therein]{lin06,vrs08,sha09}.  The flare  blast wave is believed to originate in proximity to a solar flare. This scenario is obviously based on the idea that the flare takes place explosively, like a bomb blowing out the nearby material in every direction. The earliest records of the blast wave  can be found in
the works of~~\citet{wag83} and \citet{gar84}. Recently, high-cadence EUV imaging of STEREO-EUVI and SDO-AIA has revealed in many events that the early impulsive expansion of the CME-associated magnetic structure could act as a temporary 3D piston, 
which generates a piston-driven shock that  propagates faster than the piston itself. The shock may also decouple from the piston and continue as a freely propagating shock, such a freely propagating shock is referred to as blast wave as well \citep[\textit{e.g.}, see also][and references therein]{pat09,patv09,pat10a,pat10b,war15}.
On the other hand, the CME-driven shocks are also favored by the recent  studies of coronal type II bursts \citep[\textit{e.g.},][]{manc04,manc07,lin06,ram12}.  \cite{ram12} studied forty-one metric type II radio bursts located close to the solar limb, and compared the positions of the bursts with the estimated location of the leading edge of the associated CMEs close to the Sun, their results suggest that nearly all the metric type II bursts are driven by the CME. So the origin of the metric type II burst in the solar corona is still an open question. 

 In addition to those produced by a single coronal shock, the events similar to the regular type II radio burst have also observed occasionally. These events included  the stationary type II burst due to the interaction between the reconnection outflow from the CME/flare current sheet 
with the top of flare loops or the bottom of the CME bubble \citep[\textit{e.g.},][]{aur02,aur04,aur11,man06,man09,war09,gao14a,che15}, the type-II-like continuum burst following regular type II burst as a result  of the collision of two CMEs 
 \citep{gop02,gop04}, and regular type II bursts  enhanced by the CME collision \citep{gop01,gop04,oli12}.
 Furthermore,  the  bifurcation of  the type II bursts  were observed and believed to  excite in the downstream region of the shock \citep{vrs01,man04}.  \cite{man04} also reported a bifurcation of the metric type II emission as a result of the interaction of a piston-driven shock with a vertical current sheet in the nearby helmet streamer. Recent studies reported that the broken lane of the type II bursts were caused by the shock entering the dense streamer structure from a tenuous coronal environment outside of the streamer \citep{kong12,feng12,feng13}.

In this work, we are to study a metric type II burst observed by two different spectrometers. Both data sets display an obvious broken lane on the harmonic band of the type II burst.  Associated with the radio bursts were two CMEs and flares occurring successively from the same active region, which were observed by several instruments in orbit.  The multi-band observations of the CMEs and flares shown and discussed in next section. In Section 3,  the  reasonable explanation for the formation of the broken lane structure on the type II burst will be  given; and we are discussing all results and develop conclusions in Section 4.

\section{Observations}
\label{obs}

\begin{figure}
\centering
\includegraphics[width=12.0cm]{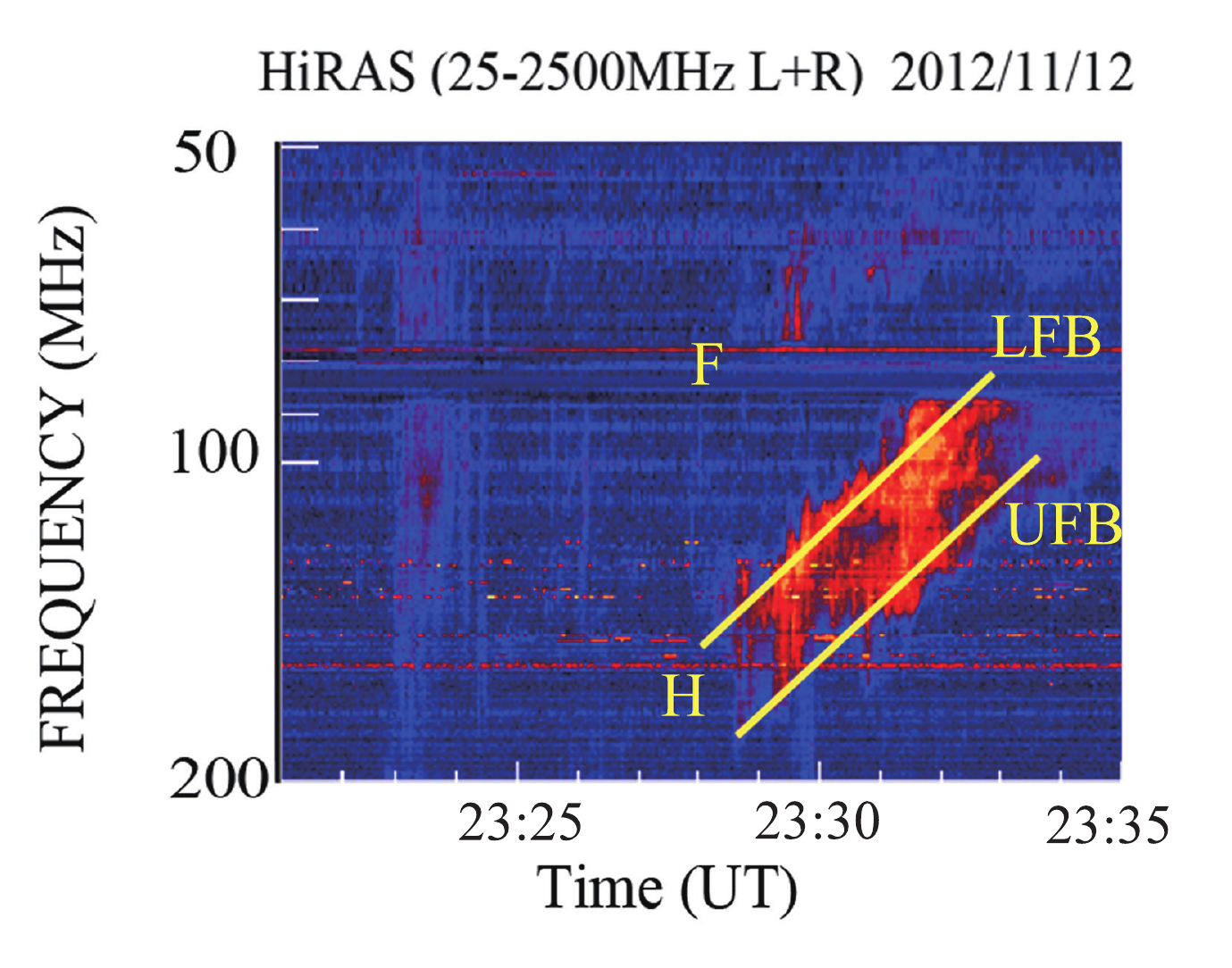}
\caption{The first type II  burst which was observed by HiRAS on November 12, 2012,  the upper and lower frequency branches (UFB and LFB, respectively) of the band-splitting on  the harmonic band indicated by the lines following the emission lanes.}
\label{fig:1}
\end{figure}

\begin{figure}
\centering

\includegraphics[width=12cm]{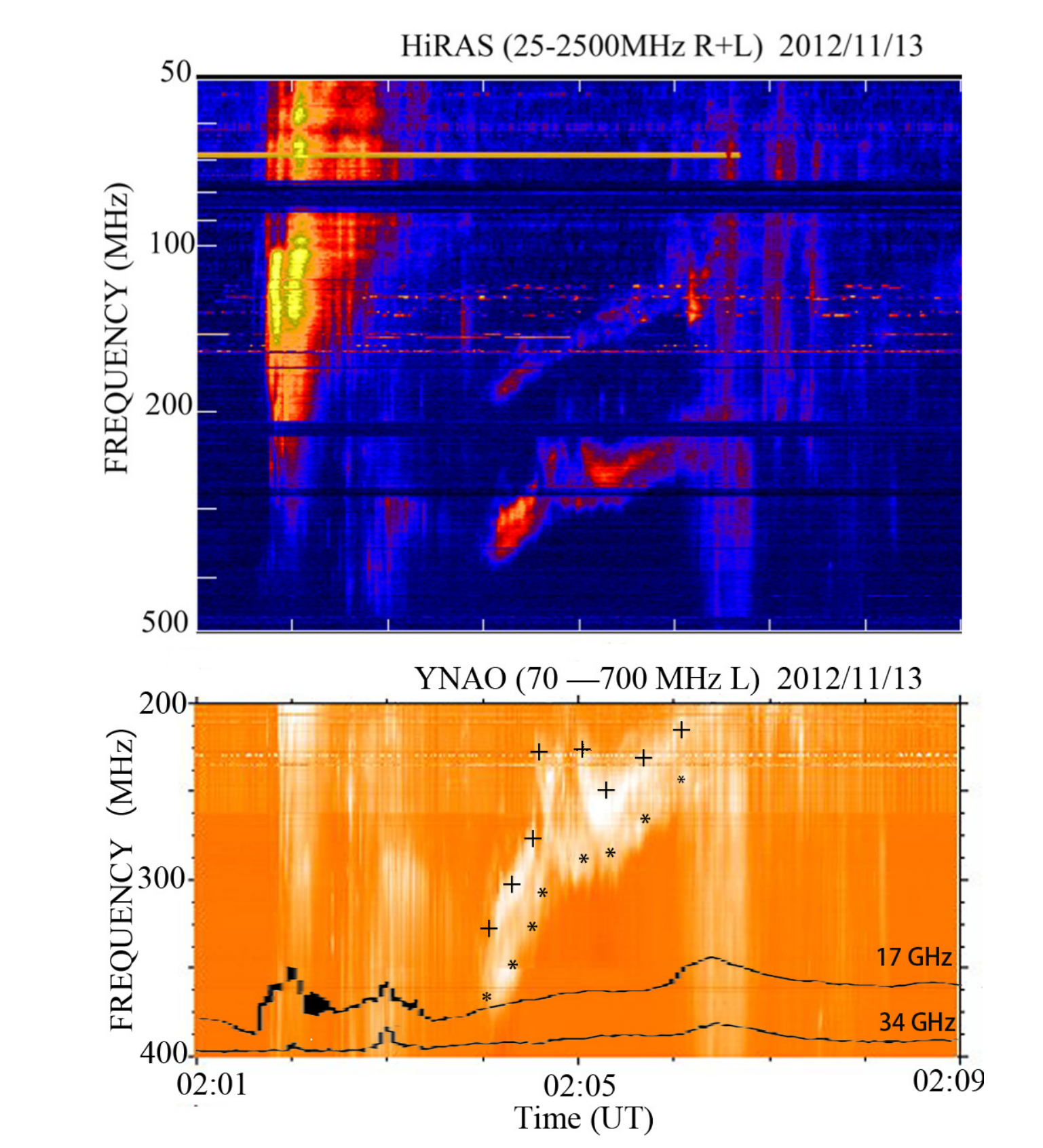}

\caption{The second type II burst was observed by the \textit{Hiraiso Radio Spectrograph}  and the YNAO spectrometer on November 13, 2012 simultaneously. The symbols `+' and `$\ast$' in bottom panel mark the lower and upper frequency  branches (LFB and UFB) of the harmonic band, respectively; two NoRH correlation plots in 17 GHz and 34 GHz  are shown  at the bottom.}  
\label{fig:2}
\end{figure}

Between 23:00 UT on November 12, 2012 and 03:00 UT on November 13, 2012, two eruptive events took place successively from the active region AR~11613. Each event produced a CME and an M-class flare. Both CMEs (or flares)  generated the type II radio burst in the metric band. The difference between these two type II radio bursts lies in that the first one displayed a pair of regular emission bands in dynamic spectrum with the harmonic band being much more significant than the fundamental one (see Figure \ref{fig:1}), while the second burst displayed a broken lane as shown in Figure \ref{fig:2}, which is more interesting to us and will be studied in detail.

The  second type II burst occurred between 02:04 and 02:09~UT on November 13, 2012, and was detected  by the \textit{Hiraiso Radio Spectrograph} \citep[HiRAS,][]{kon95} that works over the frequency range from 20 to 2500 MHz. The upper panel in Figure 2 displays the dynamic spectrum of this event obtained by HiRAS with  time cadence of  3$\sim$4~s. It shows both fundamental and harmonic band of the type II bursts. At the same time, the same event was also  observed  by the metric spectrometer  of the Yunnan Astronomical Observatories
 \citep[YNAO, see][]{gao14b}  working over the frequency range from 70 to 700~MHz with the spectral resolution of 200~kHz, the time cadence of 80~ms, as well as the high sensitivity ($<$~1~sfu). The dynamic spectrum of the event obtained by YNAO is given in the lower panel of Figure \ref{fig:2}, which shows fine structures in the harmonic band.  Two continuous curves in this panel are the \textit{Nobeyama Radioheliograph} \citep[NoRH,][]{nak94} correlation plots at 17 and 34~GHz, respectively, used as a proxy for the missing HXR data because no RHESSI data at that time interval were available. We see from the lower panel in Figure \ref{fig:2} that the three groups of type III radio bursts showing on the dynamic spectra correlated fairly well with the three peaks of the emission at both 17 and 34~GHz.

Both panels in Figure \ref{fig:2} display a clear break or gap in the harmonic band in the time interval between 02:04:25 and 02:05:15~UT, and the lower panel shows more details of fine structures because of the high frequency resolution and sensitivity of the instrument. 
Here we note that, it is difficult to recognize the break or the gap in the fundamental band of the type II radio burst from the HiRAS data (see the upper panel of Figure \ref{fig:2}), and that the YNAO spectrometer did not observe the fundamental band (see the lower panel of Figure \ref{fig:2}) because of the notch filters to suppress serious interferences in these bands  (see \citealp{gao14b} for details). So we use the harmonic band data to perform our studies in this article.

 At the same time when the above radio bursts were detected, two CMEs and the associated  solar flares were observed by several instruments as well. We have double checked the CME data in order to make sure that the CMEs and the flares observed were indeed associated with the radio bursts  considered in this paper.
 
\begin{figure}
\centerline{
\includegraphics[width=12cm]{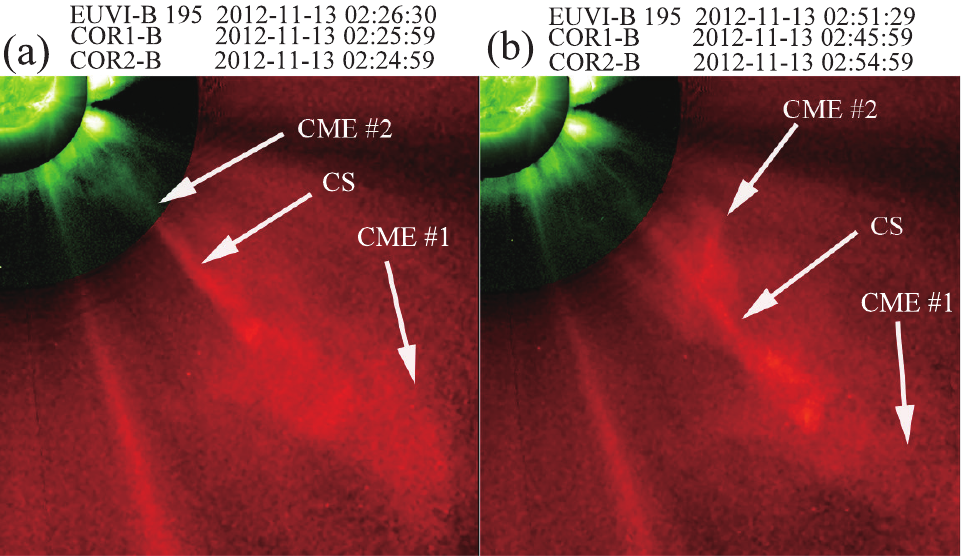}
}
\caption{Composites of EUVI-B(195 \AA)/COR1-B/COR2-B images of the CMEs at different times; the arrows in each panel specifies the CMEs front (CME \#1 and CME \#2), as well as  the current sheet (CS). }
\label{fig:3}
\end{figure}

\begin{figure}
\centering
\includegraphics[width=12cm]{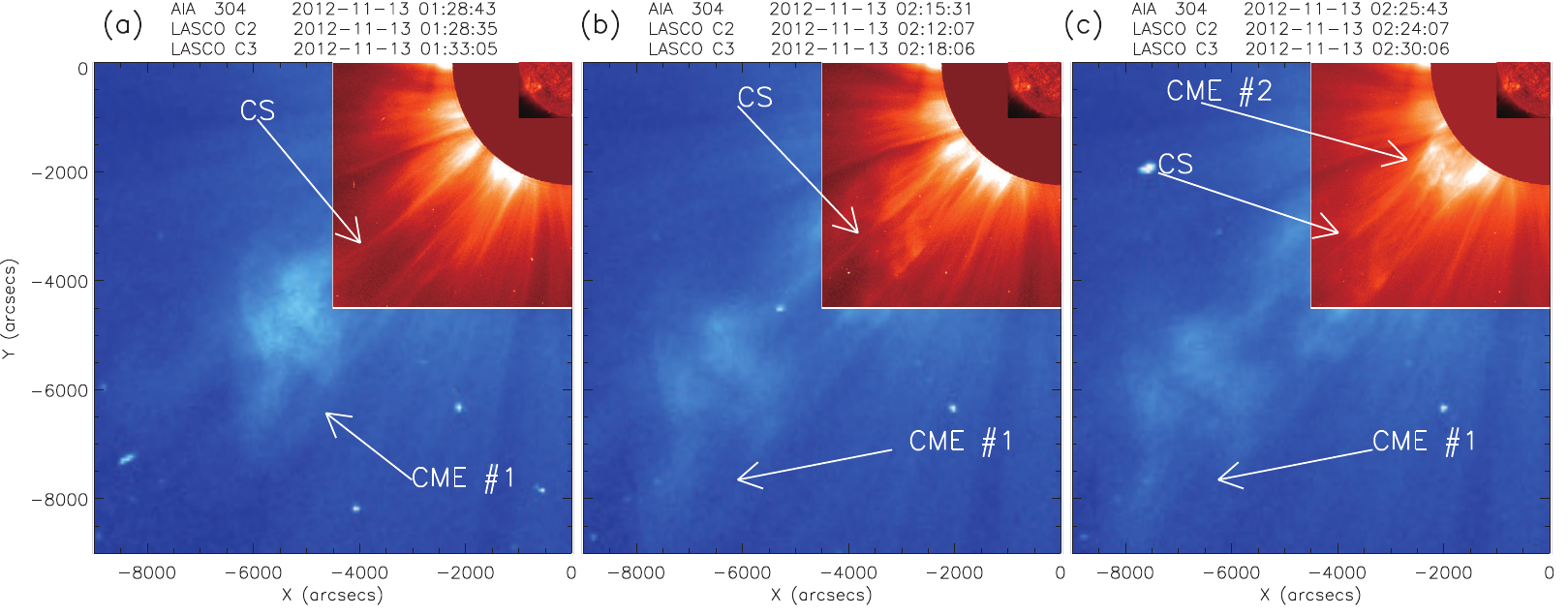}
\caption{Composite SDO/AIA--LASCO/C2--C3 images of the CMEs at different times; the arrows in each panel specify both the CME fronts (CME \#1 and CME \#2), as well as  the current sheet (CS).} 
\label{fig:4}
\end{figure}

According to the data from the \textit{Solar~~Terrestrial~~Relations~~Observatory}
\citep[STEREO,][]{kai08}, the \textit{Large~~Angle~~Spectrometric ~Coronagraph} \citep[LASCO,][]{bru95} experiment on board SOHO, and the \textit{Atmospheric Imaging Assembly} (AIA, \citealp{lem12}) on board the \textit{Solar Dynamics Observatory} (SDO), two CMEs and the  associated  flares occurred one after another, a current sheet (CS) was developed behind the first CME (see Figures \ref{fig:3} and \ref{fig:4}),  the second CME collided with the CS and the first CME successively, and the two CMEs merged eventually. The CS could be seen in the images of STEREO-B/COR1-COR2  from 00:05:59 to 02:54:59 UT on November 13, 2012.

Figure \ref{fig:3} covers the time interval between 02:24:59 and 02:54:59 UT on November 13, 2012, and displays a set of composites of EUVI-B (195 \AA)/COR1-B/COR2-B images of the two CMEs and the CS that are marked in each panel. Figures~\ref{fig:3}a--\ref{fig:3}b display the evolutionary features of the two CMEs occurred one after another, and the second CME collided with the CS, and then the second CME  caught up with the first CME. During the event, the STEREO A was located at the opposite side of the Sun, so it could not observe the event.


Figure \ref{fig:4} covers the time interval between 01:28:35 and 02:30:06 UT on
November 13, 2012, and shows a set of SDO/AIA (304 \AA)--LASCO/C2--C3 images of the CME at different times with the arrows in each panel specifying several important components of the disrupting magnetic configurations. The CS behind the first CME could be vaguely seen in Figures \ref{fig:4}a--\ref{fig:4}c. According to the sequence of various objects shown in Figures \ref{fig:3} and \ref{fig:4} together with the broken lane of type II burst shown in Figures \ref{fig:2}, we acquire a  scenario such that the two CMEs propagated roughly in the same direction, the second one moved faster than the first one, and we can infer the shock caused by the second CME (or by the second blast wave) collided with the CS at about 02:04:25 UT before the second CME caught up with the first one  (at about 04:24:59 UT). Here we note that limited by the observational data we are able to collect, we cannot determine the driver of the shock. However, the shock driver is not the main topic of this work. We will focus on the broken lane structure of the type II burst itself.

Regarding the scenario of the shock impacting on the CS described above, we need to point out that this scenario was constructed on the basis of indirect evidence since no direct evidence is available for this event. The other possible scenarios could not be ruled out. But we use the above scenario to help  analyzing the event and understanding the physics behind the observation presented in this work.

The first CME appeared in the field of view (FOV) of STEREO--B/COR1 at 23:45:59 UT for the first time on November 12, 2012 at the velocity of $\sim 700$~km~s$^{-1}$, and it was also seen at around 00:54:49 UT on November 13, 2012 by LASCO/C2--C3 at the speed of $\sim 611$~km~s$^{-1}$ according to the LASCO CME Catalog \url{(http://cdaw.gsfc.nasa.gov/CME_list)}. The difference in its speed could be influenced by projection effects because they are always only plane-of-sky speeds. The second CME was first seen by STEREO--B/COR1 at 02:25:59 UT on November 13, 2012 (Figure \ref{fig:3}a) at the velocity of $\sim 880$~km~s$^{-1}$, and it also appeared in the FOV of LASCO/C2 after 02:24:07 UT (Figure \ref{fig:4}c) at the speed of $\sim 851$~km~s$^{-1}$ according to the LASCO CME Catalog \url{(http://cdaw.gsfc.nasa.gov/CME_list)}.

\begin{figure}
\centering

\includegraphics[width=12cm]{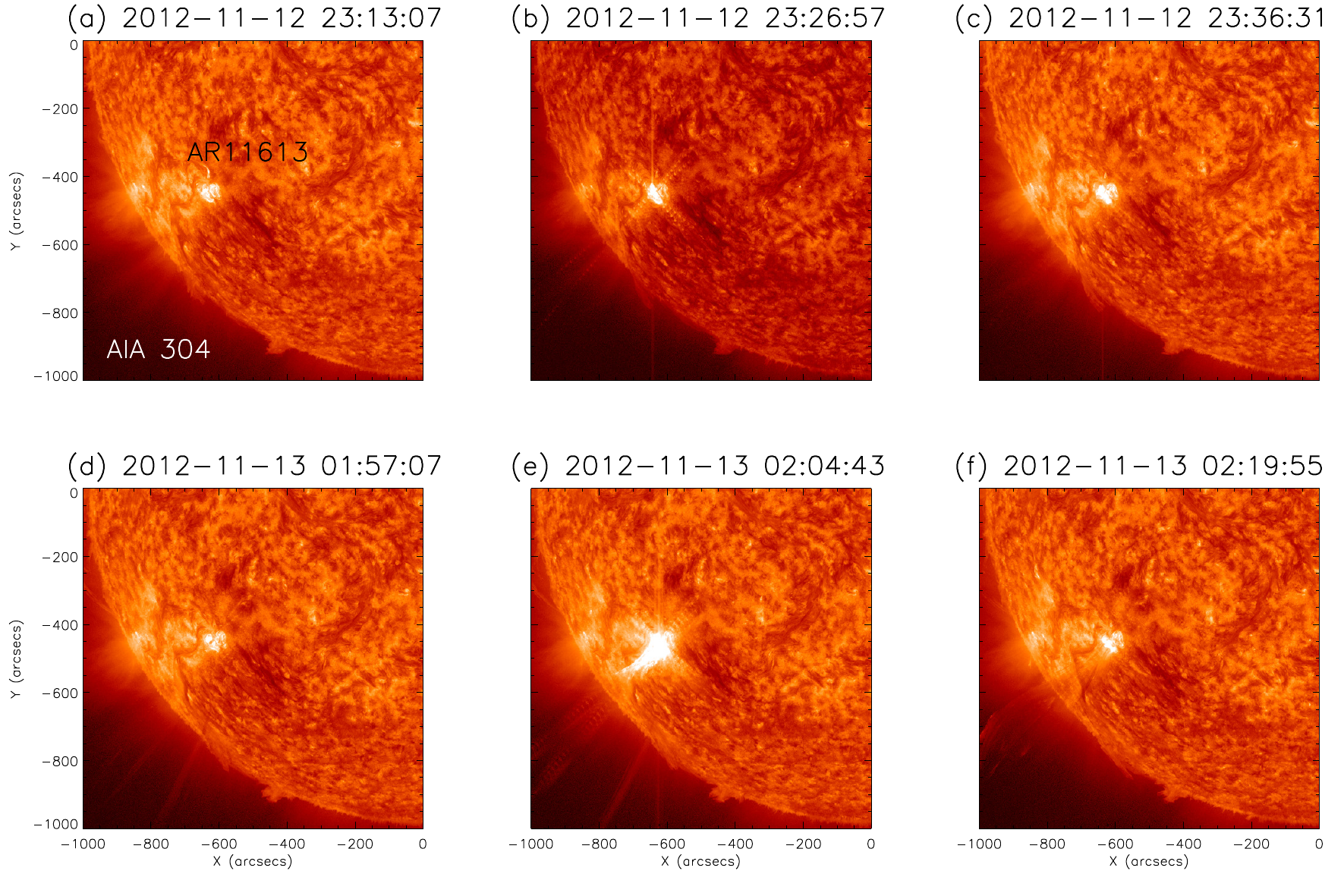}

\caption{Observational data of two eruptions. The images of SDO/AIA indicate  the  first (a--c) and second flare (d--f)  in AIA 304 \AA ~  in the active region AR 11613,  respectively.}
\label{fig:5}
\end{figure}

In addition to CME propagations, the CS could also be recognized in the STEREO B/COR1--COR2 and LASCO/C2--C3 images (see Figures \ref{fig:3} and \ref{fig:4}). Images of the CS shown in these panels indicate that the CS was observed roughly edge-on by both satellites. This allows us to measure the apparent thickness of the CS directly. From images of the STEREO B, we find that the value of this thickness is about $7\times 10^{4}$~km, as indicated by the arrow of  `CS' in Figure \ref{fig:3}a, and the LASCO data bring this value to around $4\times 10^{4}$~km as indicated by the arrow of `CS' in Figure \ref{fig:4}a, which is consistent with the results obtained for the CS apparent thickness previously in different events by various instruments in different wavelengths \citep[\textit{e.g.}, see also][]{cia13,ling14}.

Associated with the two CMEs were two flares which  occurred  in the active region
AR 11613 in succession, the first one was a GOES M2.0-class flare in soft X-rays, the onset and peak time were 23:13 UT and 23:28 UT, respectively. The second one  was a GOES M6.0-class flare, the onset and peak time were 01:58 UT and 02:04 UT, respectively. 
Figure 5 shows the time series of  two flares observed by SDO/AIA in 304  in active region AR 11613,  respectively. The active region  AR 11613 was a $\beta\gamma/\alpha\gamma$ active region, from which a total of 15 eruptions  originated on November 13, 2012.

 
\section{Data Analysis and Results} 
      \label{S-Data Analysis}      

\begin{figure}
\centering
\includegraphics[width=12cm,height=16cm]{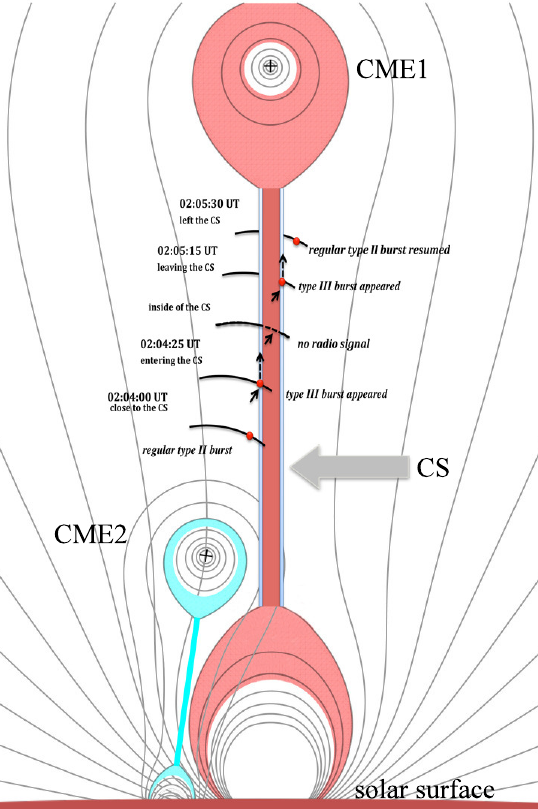}
\caption{Schematic illustration of the interaction of the CME--driven shock with a CME--flare CS developed in the previous eruption. }
\label{fig:6}
\end{figure}

According to the standard theory of  the solar eruption \citep{for96,lin00}, and the observations of white light images from STEREO-B COR1/COR2,  LASCO C2/C3 as shown in Figures \ref{fig:3} and \ref{fig:4}, we draw a cartoon (Figure \ref{fig:6}) to schematically  describe  the process of the shock  propagating through the CS.  
 As shown in the cartoon,  we  see how the fast-mode shock produced by the second CME propagated, entered and left the CS behind the first CME, as well as the expected observational consequences. The crosses indicate the center of the two CMEs, the light grey curves represent magnetic field lines, the light blue lines show the edges of the CS developed by the first CME, the red regions specify the CME bubble, flare region, and the CS of the first eruption, respectively, and the bright blue region is for the second eruption. The thick black curved lines specify the shock front,  the red dot on the shock fronts indicates the source of the type II radio burst,  the solid arrow indicates the direction of the shock propagation, the dotted curve specifies the destructed part of the shock accounting for the gap on the lane of the type II radio burst (see also the bottom panel in Figure 2), and  the dashed arrow shows the direction in which the energetic electrons around the shock escape, generating the type III radio burst that appears at the two edges of the gap.
 
  According to the scenario, a fast mode quasi-perpendicular (QPE) shock  propagated through the corona below the lower tip of the CS left by the first CME from 02:04:00 to 02:04:25 UT. Shock accelerated electrons produce the type II burst accounting for the regular `backbone' structure in the dynamic spectrum. 
When the shock moved close to the CS, on the other hand, it reached (at least) locally open magnetic field lines around the CS and turned to quasi-parallel (QPA), through which those fenced electrons escape from the shock producing the type III radio burst (drifting to lower frequency) around one edge of the CS (see Figure 2). As the  shock left the CS,  part of the shock and  the associated turbulent structures resumed,  the accelerated electrons were bound around the shock again, and the type III radio burst recovered near the another edge of the CS. Eventually, after 02:05:15 UT, the shock left the CS completely and  turned back to a QPE  one, and the type II radio burst totally recovered (see the right segment of the regular lane in the bottom panel of Figure 2).  Details and the observational consequences of this process can be seen in Figure 6 clearly.

\citet{manc04} proposed a similar scenario in their model: The strength of a MHD fast-mode shock in the corona could be very much enhanced when a shock propagates along the axes of  streamers that were identified as a typically low Alfv\'en speed structure with high density and  weak magnetic field. However, for the CME--flare CS, the situation could be different.   A CME--flare CS is usually a  high temperature region in the corona \citep{cia02, ko03,cia08,lin15}, but the electron density inside may  not be very high.  Theoretical calculations showed that the difference between the plasma density in the CS and that in the surrounding  corona
 does not exceed a factor of 4 (usually 2-3) because of the basic properties of the plasma continuity (\citealp{pri00}, p.31). The numerical experiments suggested that the density inside the sheet can be comparable to that of the surroundings (\citealp{she11}), and
the highest electron density inside the CS is about 2.2 times that of the surroundings (see Figure 11 of  \citealp{mei12}).  On the other hand, the electron acceleration efficiency strongly depends on the angle  between the
shock normal  and the upstream magnetic field. Generally a QPE
shock  favors the acceleration of electrons (\textit{e.g.}, \citealp{hol83,wu84}). Recently,  a test-particle simulation by \citet{kong16} found that the large-scale shock and the magnetic field configuration play an important role in the efficiency and the location of electron accelerations. 

Combining these pieces of knowledge with the information revealed by Figure 6, we realize that, in the event studied here, during 02:04:25 to 02:05:15 UT, inside the CS, the shock was QPA,  and less electrons could be efficiently accelerated by the shock, so the radio signal did not appear to be generated (see the gap on the broken lane of the type II radio burst in the bottom panel of Figure 2). 


The similar phenomenon of the disappearance of the radio signal was also reported by \citet{che15} recently. They noticed that a localized radio source,  as a tracer of the termination shock (TS), appeared and disappeared successively. Combining all the results they could collect for the radio burst and TS deduced from observations and numerical experiments, they concluded that the radio signal disappeared as the TS was destroyed and, as a result, number of the energetic electrons around the TS decreased dramatically (see also the Figure 2d in \citealp{che15}). Another broken lane events have been reported  by \cite{feng12}, \cite{kong12}, and \cite{che14}  for the case that the coronal shock interacted with the CS inside a nearby  helmet streamer. In the present case, on the other hand, the coronal shock collided with the post-CME CS.
 


Here we note that the crossing of other coronal structures by shocks might also result in  breaks or gaps in the type II bursts.  In this event, on the other hand, two events  occurred from the same active region  successively (see Figures \ref{fig:3}--\ref{fig:5}), and  two resultant CMEs propagated almost in the same  direction. Both CMEs (or flares)  produced type II bursts, but that produced by the first CME (or flare) did not have  broken lane in  the dynamic spectrum (see Figure \ref{fig:1}),  which indicates no plasma irregularities along the way the first CME propagated. This indirectly suggests that the broken lane of the second type II radio burst resulted from the collision of the second shock with the CS formed in the wake of the first CME.

Since plasma emission is generally considered to produce the type II emission, the lane of the type II radio burst in the dynamic spectrum gives the rate of the frequency drift, $df/dt$, of the radio emission that results from the propagation of the shock in the corona.   Relating the observed radio emission frequency $f_{obs}$ to the plasma electron frequency $f_{p}$, and then to the local electron density $n_{e}$ leads to the speed of the shock in the corona provided the dependence of $n_{e}$ on the altitude $h$ is given. Direct measurement of the slope of the type II burst lane yields the rate of the frequency drift, $df/dt$, which is about $-0.16$~MHz~s$^{-1}$ for the first shock according to Figure \ref{fig:1}, and about $-0.46$~MHz~s$^{-1}$ for the second shock according to Figure \ref{fig:2}, respectively. This may help us further deduce the altitude and the corresponding speed of the shock provided the distribution of the electron density in the corona is given.

In our calculations, the coronal electron density $n_{e}(h)=n_{0}g(h)$ is described by the empirical model of \citet{sit99}, such that
\begin{equation}
g(h)=a_{1}z^{2}(h)e^{a_{2}z(h)}[1+a_{3}z(h)+a_{4}z^{2}(h)+a_{5}z^{3}(h)],
\end{equation}
where   $z(h)= 1/(1+h),$ $ a_{1}= 0.001292$
   $a_{2}= 4.8039,$  $a_{3}= 0.29696,$
   $a_{4}= -7.1743,$ $a_{5}= 12.321,$
with~$g(0)= 1$, and $n_{0}$ = 10$^{10}$~cm$^{-3}$ being the electron density at the base of the corona.

\begin{figure}
\centering
\includegraphics[width=12cm]{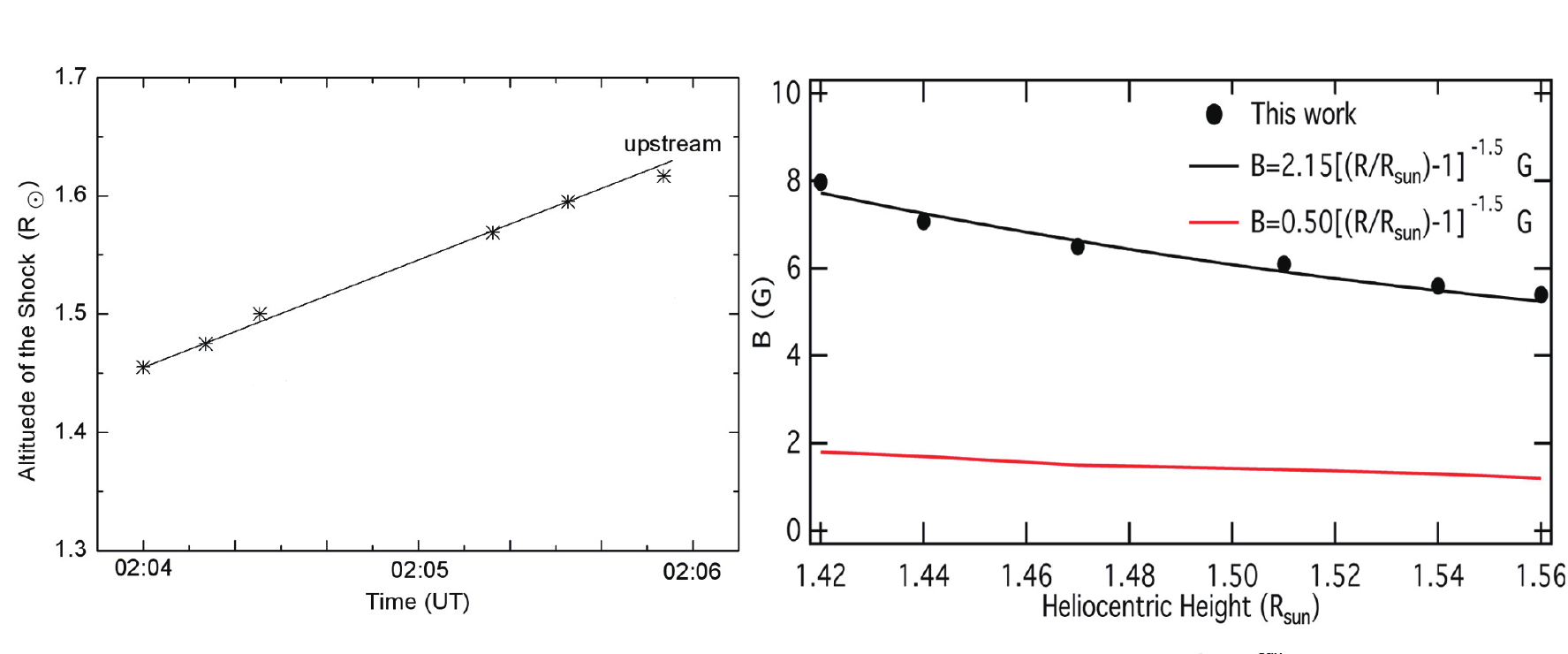}
\caption{Left panel: Temporal evolution of the altitude of  the second  shock accounting for the second type II burst. Right panel:  Strengths of the magnetic field $B$ in the upstream region of the second shock at different altitudes (solid dots), and the red and the black curves are the results calculated from the empirical model of \citet{dul78} given in Equation (\ref{eq:DM1}) and from the modified model given in Equation (\ref{eq:DM2}), respectively.}
\label{fig:7}
\end{figure}

Combining the relationship among $f_{obs}$, $f_{p}$, and $n_{e}$ with Equation (1), we obtain a speed of about 851~km~s$^{-1}$ for the first shock, according to Figure \ref{fig:1}. The changes in altitude of the second shock against time, together with the corresponding speeds, can be obtained as well according to Figure \ref{fig:2}.  The left panel of 
Figure \ref{fig:7} displays the altitude changes of the second shock, and a linear fit to these data brings the speed of the second shock to $1100$ km s$^{-1}$. We note here that the two outliers located on the spiky emission features above the type II burst in the bottom panel of Figure \ref{fig:2} at 02:04:25 and 02:05:15~UT, respectively, are used to indicate the energetic particles escaping from the shock, so unlike the other marks, the information revealed by them does not belong to the shock, but to the escaping particles, therefore, we do not plot them in Figure \ref{fig:7}.


The plots in the left panel of Figure \ref{fig:2} also indicate that the downstream of the shock is perturbed more apparently than the upstream as the shock passed through the CS region. This is probably because the magnetized plasma in the downstream region is more turbulent than that near the upstream, which causes more diffusion and thermalization of energetic particles, leading to weakening of radio emission in the downstream region. Looking into the gap on the type II burst lane displayed in Figure \ref{fig:2}, we realize that the time interval covered by the gap is about 50 s (from about 02:04:25 to 02:05:15~UT). Multiplying the time interval of 50 s and the speed of the shock upstream gives a scale of the gap in space, which is $5.5\times 10^{4}$~km. This value represents a kind of extension of the CS region in space, but may not be the thickness of the CS. Several uncertainties, including those in electron density and inclination of the shock front to the CS, make it difficult to relate this value to the thickness of the CS although it is close to the apparent value of the the CS thickness obtained earlier \citep[\textit{e.g.},][]{lin05,lin09,cia08,vrs09,sav10}.

In addition, if  the shock was driven by the second CME as a bow shock,  we can also estimate the strength of the magnetic field at the location,  where the type II radio burst was just initiated as the CME speed started to exceed the local speed of the fast magneto-acoustic wave, which is approximately the local Alfv\'en speed in the low corona, and a shock commenced to form in front of the CME. The dynamic spectra shown in Figure \ref{fig:2} give the start frequency of the type II radio burst, and then the corresponding electron density, $n_{e}$ according to relations of $n_{e}$ to $f_{obs}$ and $f_{p}$. Equating either the CME speed or the shock speed deduced from the dynamic spectrum for a given model of $n_{e}$ to the Alfv\'{e}n speed, $v_{A}=B/\sqrt{4\pi m_{p}n_{e}}$, where $m_{p}$ is the mass of the proton, we are able to estimate the magnetic field, $B$. In the present case, the speed of the second CME was 851~km~s$^{-1}$, and the shock speed was about 1100~km~s$^{-1}$. 
Eventually, we obtained that the magnetic field strength was 7.5 G at the altitude of about 
1.46~R$_{\odot}$. Similarly, the information we just deduced for the first shock helps us deduce the strength of the magnetic field was 2.6 G at the altitude of about 1.76~R$_{\odot}$.  However, a CME may also generate a piston-driven shock, for which it is not necessary that the driver exceeds the local fast-mode speed (see the discussions of \citealp{vrs08}, and \citealp{war15}). So the fact that 
CME is some 25\% slower than local fast-mode speed  may not be due to the density model or uncertainties in the starting location of the type II burst, but due to the decoupling of the shock from the driver. Therefore taking into account this extra 25\% increment to the shock speed we obtained that the magnetic field was 9.4 G at the altitude of about 1.46~R$_{\odot}$ for the second shock.  For the first shock, the magnetic field was  3.3 G at the altitude of about 1.76~R$_{\odot}$.


On the other hand, we are able to estimate the magnetic field of the source region of the type II radio burst based on the band-splitting of  the first  and second type II burst (Figures 1 and 2). As a result of the plasma emission from the upstream and downstream shock regions, the band-splitting frequencies indicate the electron densities behind and ahead
of the shock front \citep[see the detailed discussions of][]{vrs01,vrs02,vrs04}. So the band-splitting width of the first type II burst revealed the density jump across the shock wave, and the Rankine-Hugoniot (RH) relation for the shock wave can be used to deduce the related parameters \citep[see also][]{sme75,man95,pri00,vrs02,cho07}. The relationship between the downstream and upstream density jump $X$ (compression) and the Alfv\'en Mach number $M_{A}$ depends on the plasma $\beta$ (the ratio of the gas pressure to the magnetic compression) and  the angle $\theta$ between the shock normal and the upstream magnetic field \citep[see also][]{pri00,vrs02}. Here, $M_{A}$ is the speed $v_{sh}$ of the shock front compared to the local Alfv\'en speed $v_{A}$,  $M_{A}=v_{sh}/v_{A}$. The relative band-split (band distance width, BDW) is defined as: 
\begin{equation}
BDW= \Delta f/f_{l}=(f_{u}-f_{l})/f_{l},
\end{equation}
where $f_{u}$ and $f_{l}$ are the frequencies measured at the upper and the lower frequency branches as UFB and LFB (see Figure 1)  respectively. The density jump, $X$, across the shock is defined as:
\begin{equation}
X = N_{2}/N_{1} = (\frac{f_{u}}{f_{l}})^{2}=(BDW+1)^{2},
\end{equation}
where $N_{1}$ and $N_{2}$ are the electron densities upstream and downstream of the shock, respectively.  $f_{u}$ and $f_{l}$ could be measured directly from UFB and LFB, and deduce $X$ from Equation (3). Furthermore, $X$ is related to $M_{A}$ on the basis of the standard theory of the the fast-mode shock (\textit{e.g.}, see  \citealp{pri00}, p. 31). In the case of the perpendicular shock ($\theta$ = $90^{\circ}$), with $\beta \rightarrow 0$ and $\gamma=5/3$ in the corona \citep{vrs02} ,  $M_{A}$ is related to $X$ in the way of
\begin{equation}
M_{A}=\sqrt{\frac{X(X+5)}{2(4-X)}}.
\end{equation}

To this point, we are able to deduce several important parameters mentioned above for the first type II radio burst. Table 1 lists $X$ and $M_{A}$, together with the other parameters at different times. We see that in the time interval of the first type II burst from 23:28 to 23:33 UT, values of  $BDW$ varied from 0.29 to 0.33, which corresponded to the density jump ($X$) between 1.67 and 1.77, and the Alfv\'en Mach number ($M_{A}$) ranging from 1.54 to 1.64. The mean values of these three parameters are 0.31, 1.73, and 1.60, respectively.

With the speed of the first shock being known, $v_{sh}=851$~km~s$^{-1}$, and the Alfv\'{e}n Mach number calculated above, we can obtain $v_{A}$ is between 520 and 560~km~s$^{-1}$ within the height range from 1.7 to 2~R$_{\odot}$. Then, the ambient magnetic field strengths in the way the shock propagated is deduced from $B_{G}=v_{A}\sqrt{4\pi m_{p} n_{e}}$, ~$n_{e} [cm^{-3}]=({f_{l} [kHz]}/{8.98})^{2}$, according  to the information  revealed by  the LFB. This eventually brings the electron number density to the range from  6 to 1$\times$10$^{7}$~cm$^{-3}$, and the magnetic field to the range from  2 G at 1.7~R$_{\odot}$ to 1 G  at  2~R$_{\odot}$, respectively.

\begin{table}
\caption{Values of $f_{u}(t_{i})$ and $f_{l}(t_{i})$ at the upper and lower frequency branches (UFB and LFB, respectively) of the band-splitting on the harmonic band in the type II burst November 12, 2012 and deduced shock parameters}

\label{T-simple}
{
\begin{tabular}{clccccccc}     
  \hline                   
Time      &  &  UFB  & LFB  &$X$&$BDW$ & $M_{A}$\\
(UT)      &  & (MHz) & (MHz)&      &               &            \\ 
  \hline
23:28    &  & 192& 149   &  1.67     & 0.29  &  1.54\\ 
23:29    &  & 175& 132   &1.77       & 0.33  &  1.64 \\
23:30   &  & 153& 116     &   1.74  &   0.32 &   1.61\\ 
23:31    &  & 136   & 104  &  1.72    &   0.31 &  1.59\\ 
23:32   &  & 121 & 92       &    1.74   &   0.32  &  1.61\\ 
23:33   &  & 106  & 81      &    1.72 &   0.31 &    1.59 \\ 

\hline
\end{tabular}
}
\end{table}

\begin{table}
\caption{Values of $f_{u}(t_{i})$ and $f_{l}(t_{i})$ at the upper and lower frequency branches (UFB and LFB, respectively) of the harmonic band in the type II burst November 13, 2012 and deduced shock parameters
}
\label{T-simple}
{
\begin{tabular}{clccccc}     
  \hline                   
Time      &  &  UFB  & LFB  &$X$&$BDW$ & $M_{A}$  \\
(UT)      &  & (MHz) & (MHz)& &       &    \\
  \hline
02:04:00    &  & 368& 325   &1.28& 0.13 & 1.22\\
02:04:15  &  & 350 & 300   &1.37& 0.17 & 1.29\\
02:04:25  &  & 325  & 275 &1.39& 0.18    & 1.30 \\
02:05:15  &  & 285  & 250  &1.3& 0.14  & 1.23 \\
02:05:30  &  & 263& 230 &1.3& 0.14  & 1.23 \\
02:05:53  &  & 250  & 220   &1.3 & 0.14   & 1.23 \\

\hline
\end{tabular}
}
\end{table}



Similarly, for the second type II burst with the broken lane, we could also estimate the magnetic field of the source region of the type II radio burst according to the feature and fine structures of the signal lane as shown in Figure \ref{fig:2}. The data of the YNAO spectrometer with very high frequency resolution ($\sim 200$ kHz) and sensitivity (see \citealp{gao14b} for more details) allow us to perform this investigation.  Table 2 lists $X$ and $M_{A}$ at different time. We can see that  the type II radio bursts covered a time interval  from 02:04:00 to 02:05:53 UT. The value of $BDW$ varies from 0.13 to 0.18, which leads to a density jump ($X$) between 1.28 and 1.39, and the Alfv\'en Mach number ($M_{A}$) ranging from 1.22 to 1.30. The mean values of them are $\langle BDW \rangle$= 0.15, $\langle X \rangle$ =1.32, and  $\langle M_{A} \rangle$ =1.25, respectively. Furthermore, as the shock crossed the CS from 02:04:25 to 02:05:15 UT,  the $BDW$ was within the range from  0.18 to 0.14,  which leads to a density jump ($X$) varying from 1.39 to 1.3, and an Alfv\'{e}n Mach number ($M_{A}$)  from 1.30 to 1.23. 


 The altitude of the second type II burst \textit{versus} time as shown in left panel of Figure \ref{fig:7}, the shock speed $v_{sh}$ is about 1100 km s$^{-1}$, and thus the $v_{A}$ is  between 840 and 900~km~s$^{-1}$.
 Bringing the information we could have collected together, we are able to obtain the magnetic field at different height and different time as shown in the right panel of Figure \ref{fig:7}. The result indicates that the electron density was within the range from  3.7 to 1.7$\times$10$^{8}$ cm$^{-3}$, 
from which we deduced the magnetic field strengths decreasing from 8~G at 1.42~R$_{\odot}$ to 5.4~G at 1.56~R$_{\odot}$ (see black dots in the right panel of Figure 7). This result could be compared to that given by the empirical model of \citet{dul78}:
\begin{equation}
B=0.50[(R/R_{\odot})-1]^{-1.5}, \label{eq:DM1}
\end{equation}
where $R$ is the heliocentric height, and $B$ is in G. Plotting the results for $B$ directly calculated from Equation (\ref{eq:DM1}) gives the red curve, from which we noticed that the values of $B$ deduced from the observations in this work are about 6~G larger than those calculated from Equation (\ref{eq:DM1}). Carefully adjusting the factor in front of parenthesis from 0.50 to 2.15, say
\begin{equation}
B=2.15[(R/R_{\odot})-1]^{-1.5}, \label{eq:DM2}
\end{equation}
and recalculating the values of $B$ in the same height interval, we obtain the black curve in Figure 7, which is nicely fit to the results deduced for the event studied in this work.

In addition, the magnetic field could also be deduced through an alternative approach \citep[\textit{e.g.},][]{war05,cho07,she12,xue14}. Using the electron number density n$_{e}$ $\approx$ 10$^{8}$~cm$^{−3}$ estimated on the basis of the SDO/AIA in 171 \AA~data, \cite{she12} calculated  the magnetic field $B$ of the open active region coronal loops, and they obtained $B\approx 4.24$~G.  In the other cases, \citet{vrs09} also found the ambient magnetic field was about 0.9--1.7~G for the values of electron number density between $5.0\times 10^{7}$ and $1.0\times 10^{8}$~cm$^{-3}$.


\section{Conclusions} 
      \label{S-conclusions}      

Two eruptive events took place successively in the active region AR 11613 from 23:00 UT on  November 12, 2012 to 03:00 UT on November 13, 2012. Each event produced a CME and an M-class flare.  The resultant type II radio burst produced in the first event was observed by the spectrometer of HiRAS, and that created in the second event was observed by the spectrometer of HiRAS and YNAO simultaneously. The first type II radio burst was a regular one with the band-splitting structure in both the fundamental and harmonic bands with the signal at the fundamental band less clear; the  second one displayed an interesting broken lane on the harmonic band. Since the second one was detected by both the dynamic spectrometers at YNAO and HiRAS  at 02:04 UT on November 13, 2012, we are able to comprehensively investigate the causes of the broken lane on the harmonic band of the  second type II burst.


Observations showed that the first CME left  a long CS behind it, the second CME propagated fast, then the broken lane on the harmonic band of the second type II  burst was generated. So we suggest that the fast shock was generated by the second CME or the second  blast wave,  that produced the second type II radio burst with the frequency drifting rate of --0.46 MHz~s$^{-1}$ and the onset frequency of about 308 MHz for harmonic band. Soon after its formation,  the shock swept the CS, which caused the broken lane of the type II radio burst observed by both YNAO and HiRAS dynamic spectrometers.  The broken lane of the type II radio burst as a result of the collision of the coronal shock with the CS inside the helmet streamer has been reported previously \citep[\textit{e.g.},][]{kong12,feng12,feng13,she13}. It is the first time, to our knowledge, that the collision between a shock with a post-CME CS may have been observed.  We draw this conclusion because the first type II radio burst did not show any abnormal feature, which indicates no unusual or irregular structure existing in the way of the associated shock propagating, and the broken lane of the second type II radio burst occurred when the shock that was responsible for the second type II burst just went through the CS left behind the first CME (see also Figures \ref{fig:2} through \ref{fig:4}).

 We note here that no direct evidence for the shock impacting on the CS was obtained in this event. What we have instead is the indirect evidence that could be used to deduce a certain scenario that is consistent with observations. So we are not able to rule out the other possible scenario that might also be consistent with observations. This is an open question and we need to look into it in detail in the future.



From the Figures \ref{fig:3} and \ref{fig:4}, we  estimated the width of  CS  is about 
(4--7) $\times$10$^{4}$~km. On the other side, the thickness of  CS also can be calculated according to the gap width of the broken lane of the type II radio burst resulted from the shock propagating through the  partly open magnetic field near the CS. The time of the shock took to pass through the gap is about 50 s,  the shock velocity of about 1100 km s$^{-1}$,  so the distance of the shock moving in the CS is about 5.5$\times$10$^{4}$ km. Considering the angle between the shock normal and the CS normal   is unknown, we note that the value should be an upper limit of the CS thickness. Both values obtained here are consistent with  the results obtained for the CS apparent thickness by the other authors previously (\textit{e.g.,} \citealp{lin05,lin07,lin09,lin15,cia08,vrs09,sav10,cia13}).

In addition, the magnetic field strength in the source region of each type II burst was also deduced in different ways.  First, if we assumed that each shock was driven by CME as the bow shock, the magnetic field was deduced by equating the Alfv\'{e}n speed with the speed of each CME at the moment when the  type II radio burst was generated as a result of the ignition of the CME-driven shock, and we found that the magnetic field  strength was 7.5 G at about 
1.46~R$_{\odot}$ and 2.6 G at about 1.76~R$_{\odot}$, respectively. Alternatively, if the  shock was driven by the CME as a piston, we  obtained that the magnetic field was 9.4 G at about 1.46~R$_{\odot}$  and  3.3 G at about 1.76~R$_{\odot}$, respectively. Second, the density jump, $X$, across the shock, and then the Alfv\'{e}n Mach number, $M_{A}$, of the shock, could be estimated from the band-splitting of the type II burst; multiplying $M_{A}$ with the shock speed gives the local Alfv\'{e}n speed, as well as the local magnetic field with a given density model of the corona. Eventually, we found that for the first type II burst, the magnetic field was between  1 and 2~G in the height range from 1.7 to 2 R$_{\odot}$; and for the second type II burst with the broken lane, the magnetic field was between  5.4 and 8~G in the height range from 1.42 to 1.56~R$_{\odot}$ (see the right panel of Figure \ref{fig:7}).

 On the issue of measuring the thickness of the CME/flare CS, popular approaches are through analyzing white-light images, the spectral data, and filtergrams in wavelengths suitable for plasma  diagnostics  of the CS (see the recent review by \citealp{lin15} for a detailed discussion). In this work, on the other hand, we performed plasma diagnostics and estimated the thickness of the CS through studies of the radio data with high temporal and spectral resolutions, and obtained the results that are consistent with those obtained via the other approaches. This indicates that the radio observation of high spectral and temporal resolution is a very important and valuable supplement to the other methods for investigating the CME/flare CS.
 
 By analyzing the data from the YNAO radio spectrometer, which possesses very high spectral resolution and sensitivity, we investigated the process in which a  shock approached and accessed  a CME/flare CS as well as the consequences. Our results indicated that the interaction  between the shock and the CS, the QPE shock turned to the QPA shock yielding the leak of the energetic electrons bound around the shock and resulting in the type III radio burst near the CS and the disappearance of  the type II radio burst temporarily.


{\bf Acknowledgements}

This work was supported  by Program  973~grant
 2013CBA01503,
NSFC grants U1431113, 11273055, 11333007, 11173010, 10978006 and 11403099, and CAS grants XDB9040202 and QYZDJ-SSW-SLH012. B. T.  acknowledges supports
from NSFC grants 11273030, 11373039 and 11433006. We also acknowledge
the supports of CAS ``Light of West China" Program and  Key Laboratory of Solar Activity (KLSA201405), National Astronomical Observatories of China CAS. 

{\bf Disclosure of Potential Conflicts of Interest} The authors declare that they have no conflicts of interest.


\end{article} 

\end{document}